\documentclass[twocolumn,msam,noshowpacs,superscriptaddress,preprintnumbers,amsmath,amssymb,nofootinbib]{revtex4}
\usepackage{verbatim}
\usepackage{natbib}
\usepackage[dvips]{graphicx}
\usepackage{xspace}
\usepackage{color}
\usepackage{array}
\usepackage{subfig}
\usepackage[english]{babel}
\usepackage{caption}
\usepackage{hyperref}

\newcommand{\ie}{\textit{i.\,e.}\@\xspace}
\newcommand{\etc}{etc.\@\xspace}

\newcommand{\phii}{II. Physikalisches Institut, Universit\"{a}t zu K\"{o}ln, 
	Z\"{u}lpicher Stra{\ss}e 77, D-50937 K\"{o}ln, Germany}
\newcommand{\phE}{Sincrotrone Trieste S.C.p.A., Strada Statale 14-km 163,5 
	in AREA Science Park, 34149 Basovizza, Trieste, Italy}
\newcommand{\phD}{Max Planck Institute for Chemical Physics of Solids, 
	N\"{o}thnitzer Stra{\ss}e 40,
	01187 Dresden, Germany}
\newcommand{\phB}{Department of Physics, University of Trieste, 
	Via A. Valerio 2, 34127 Trieste, Italy}
\newcommand{\ycto}{Y$_{0.63}$Ca$_{0.37}$TiO$_3$\@\xspace}
\begin{document}
\title{Temperature dependent percolation mechanism for conductivity in Y$_{0.63}$Ca$_{0.37}$TiO$_3$ 
	revealed by a microstructure study}
\author{R. German}
\affiliation{\phii}
\author{B. Zimmer}
\affiliation{\phii}
\author{T. C. Koethe}
  \affiliation{\phii}
\author{A. Barinov}
  \affiliation{\phE}
\author{A. C. Komarek}
\affiliation{\phii}
\affiliation{\phD}
  \author{M. Braden}
  \affiliation{\phii}
\author{F. Parmigiani}
\affiliation{\phii}
\affiliation{\phE}
\affiliation{\phB}
\author{P.H.M. van Loosdrecht}
  \affiliation{\phii}

\date{\today}
\hyphenation{Raman}

\begin{abstract}

\noindent We have performed optical microscopy, micro-photoelectron spectroscopy,
and micro-Raman scattering measurements on Y$_{0.63}$Ca$_{0.37}$TiO$_3$ single 
crystals in order to clarify the interplay between the microstructure and the 
temperature dependent electronic transport mechanisms in this material. 
Optical microscopy observations reveal dark and bright domain patterns 
on the surface with length scales of the order of several 
to a hundred micrometers showing a pronounced temperature 
dependent evolution. Spatially resolved photoelectron spectroscopy 
measurements show the different electronic character of these domains.
Using micro-Raman spectroscopy, we observe a 
distinct temperature dependence of the crystal structure of these domains.
On the basis of these findings the different domains are assigned to 
insulating and metallic volume fractions, respectively.  
By decreasing the temperature, the volume fraction of the conducting 
domains increases, hence allowing the electrons to percolate through 
the sample at temperatures lower than $\sim$150 K.
\end{abstract}

\maketitle


\section{Introduction}

In the past decades the perovskite YTiO$_3$ has attained much interest 
as a toy material for correlated electron systems due to its 
simple Ti-$3d^1$ electronic structure.\cite{Imada1998, Yee2015, Cao2015, Yang2017} 
It is classified as a typical Mott-Hubbard-insulator with a Mott-Hubbard 
gap of $\sim 1$ eV and charge transfer gap of  $\sim$ 4 eV.\cite{Katsufuji1995,Okimoto1995,Morikawa1996}
By substituting Y$^{3+}$ with Ca$^{2+}$ in 
Y$_{1-x}$Ca$_{x}$TiO$_3$, \ie hole-doping, first the
ferromagnetic order
vanishes at $x\sim 0.15$. 
By further doping, the macroscopic electronic property of 
this material changes
from insulating to metallic at $x > 0.39$.
For doping between $x=0.33$ and $x=0.39$ these 
samples exhibit a peculiar 
temperature driven metal-insulator transition (MIT), 
meaning that they are conducting at low temperatures and become 
insulating at higher temperatures with a transition 
temperature ranging from $\sim 90$ K to $\sim 220$ K with 
increasing $x$.\cite{Taguchi1993,Tokura1993} 
This is rather uncommon as mostly the insulating phase is found 
to be more stable at lower temperatures. The most 
remarkable in the doping range of $x = 0.33 - 0.39$ is the 
compound  Y$_{0.63}$Ca$_{0.37}$TiO$_3$, which exhibits in 
addition a hysteresis in the electrical 
resistivity and the magnetic susceptibility.\cite{Iga1996}

Y$_{0.63}$Ca$_{0.37}$TiO$_3$ undergoes a structural 
phase transition from high temperature orthorhombic 
Pbnm (HTO) to low-temperature monoclinic P2$_1$/n (LTM) 
symmetry at $\sim$200 K.\cite{Kato2002} 
This LTM phase has been shown to be orbitally ordered,
and possibly also charge ordered.\cite{Nakao2004}
As seen in electrical resistivity measurements, 
the compound is insulating at room temperature and the 
MIT for this compound takes place at $\sim 150$ K.\cite{Tsubota2003}
\footnote{The lattice parameters and physical properties 
of our samples with the chemical composition x = 37 \%, as determined experimentally 
(see Sec. II), correspond to samples with x = 39 \% in ref. \cite{Tsubota2003}.} 
Thus the crystallographic transition from HTO to 
LTM is not related to a change in the macroscopic electronic 
property and the sample remains insulating in the LTM phase.
Upon further cooling below 200 K, a low-temperature orthorhombic
Pbnm (LTO) phase appears and starts to coexist with the LTM phase.\cite{Kato2002} 
This LTO phase is distinguished from the HTO phase, 
as it has a smaller unit cell than the HTO phase.\cite{Nakao2004}  
The volume fraction of the LTO phase increases 
by lowering the temperature, accompanying the 
metal insulator transition.\cite{Tsubota2003} Thus, a possible 
conducting character of the LTO phase would suggest a percolation 
mechanism underlying the MIT.
The phase separation is also observed in 
conducting samples $(x > 0.40)$, where the volume fraction
of the LTO phase is high enough even at room temperature to 
render these samples conducting, assuming the LTO phase to be 
metallic.\cite{Matsuhata2004}
\linebreak
In this work we will directly show the relationship of the 
crystallographic structure of Y$_{1-x}$Ca$_{x}$TiO$_3$ and 
its macroscopic electronic properties. For this we 
investigated the material Y$_{1-x}$Ca$_{x}$TiO$_3$ 
with $x = 0.37$, exhibiting the pronounced hysteresis 
at the MIT.\cite{KomarekPhD} Optical microscopy studies 
allow to follow the temperature and spatial evolution of 
coexisting domains in this material. Using micro photoemission 
studies it is possible to classify the coexisting domains to 
conducting and insulating phases. Finally, using micro Raman 
studies, these coexisting domains are also classified to 
their crystallographic structures, hence allowing to directly 
relate the microstructure to the electronic behavior of this 
material, underlying the temperature driven MIT.

\begin{figure*}[tbh]
	\subfloat[ T = 10 K; (001)]{\includegraphics[width=.3\textwidth]
		{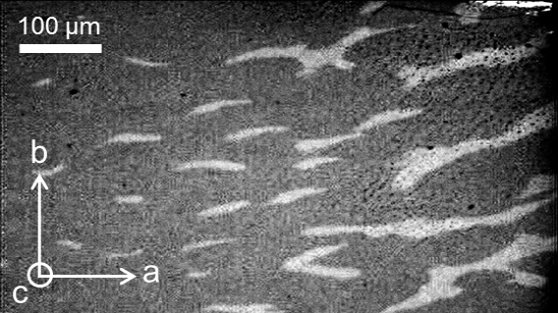}\label{sampleC}}\vspace{1mm}
	\subfloat[ T = 10 K; (010)]{\includegraphics[width=.3\textwidth]
		{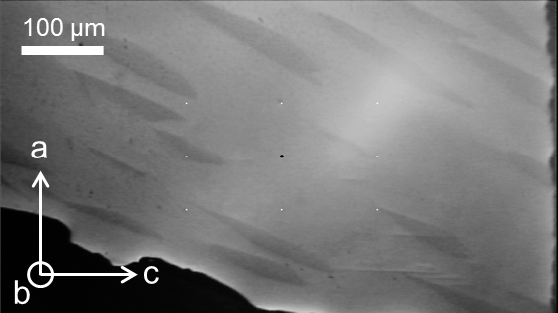}\label{sampleB}}\vspace{1mm}
	\subfloat[ T = 10 K; (100)]{\includegraphics[width=.3\textwidth]
		{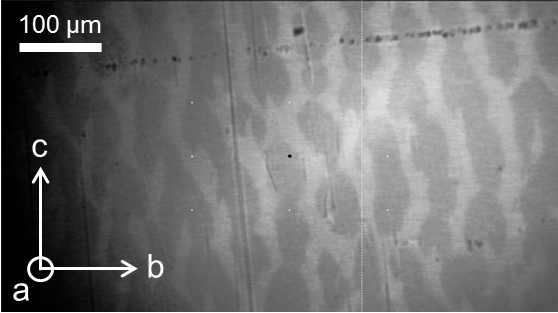}\label{sampleA}}\\
	\subfloat[T = 100 K; (001)]{\includegraphics[width=.3\textwidth]
		{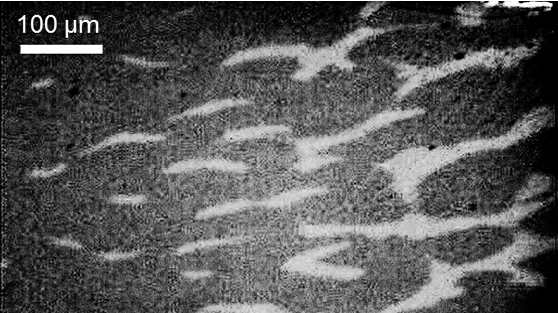}\label{sampleC_100}}\vspace{1mm}
	\subfloat[T = 170 K; (001)]{\includegraphics[width=.3\textwidth]
		{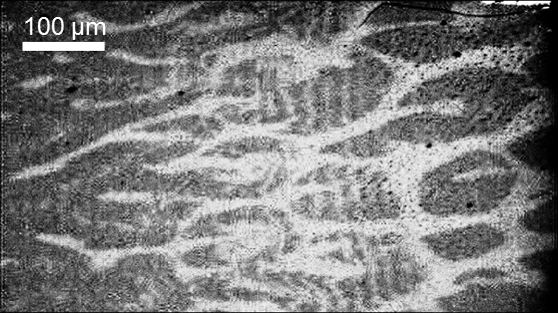}\label{sampleC_170}}\vspace{1mm}
	\subfloat[T = 296 K; (001)]{\includegraphics[width=.30\textwidth]
		{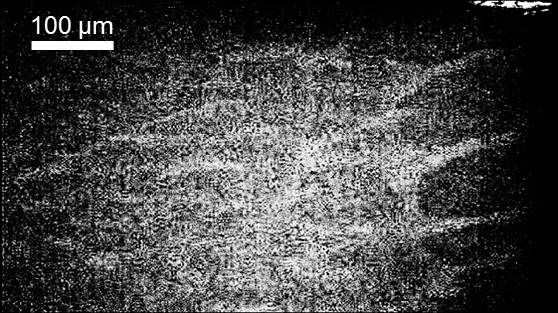}\label{sampleC_296}}
\captionsetup{justification=raggedright,singlelinecheck=false}

\caption{Microscope images of \ycto. (\ref{sampleC}) - (\ref{sampleA}) 
	pattern formation for the three surface orientations, 
	(001), (010), and (100), respectively, at 10K. (\ref{sampleC_100}) - 
	(\ref{sampleC_296}) temperature evolution of the (001)-surface pattern.}
	\label{microscope}
\end{figure*}

\section{Experimental}
The untwinned single crystal Y$_{0.63}$Ca$_{0.37}$TiO$_3$ was grown 
by the floating-zone technique. 
The oxygen stoichiometry was checked by thermal gravimetric analysis (TGA) to be $3\pm0.02$, while the Ca content was determined from a comparison
of the measured lattice constants with the results of ref.\cite{Tsubota2005} (see also ref.\cite{KomarekPhD}, figure 5.10) to be $x=0.37\pm0.01$.
The optical microscopy studies of the 
mechanically polished samples were done in the back-reflection geometry, 
using a microscope with an objective of 8x magnification and numerical aperture of 0.18.
The photoemission microscopy was performed at the spectromicroscopy beamline operated by 
the Elettra storage ring.\cite{Barbo2000} Photons at 74 eV were focused through a 
Schwarzschild objective, to obtain a submicron size spot on the sample. The measurements
were performed in ultra-high vacuum at a base pressure of $p<2\times10^{-10}$mbar on 
$in situ$ cleaved (001)-surfaces. The temperature of the sample was  T $=110$ K throughout 
the photoemission measurements.

For the spontaneous micro-Raman setup a continuous wave DPSS laser of 532 nm wavelength 
was used as a light source, sent through a microscope with a long distance objective and 
a numerical aperture of 0.2. The Raman scattered light was collected in a back-scattering 
geometry using a triple subtractive spectrometer equipped with a liquid nitrogen 
cooled CCD camera. This setup yields an energy resolution of 7 cm$^{-1}$. The samples 
were mounted in an Oxford Microstat cryostat, cooled by liquid Helium.

In this work the following back-scattering geometry for the Raman measurements was used: 
the incoming laser light propagates in the $c$-direction of the sample and the polarization 
of the incident and detected scattered light were each directed in the $a$-direction of 
the sample, abbreviated in the Porto notation as $c(aa)\bar{c}$. With the microscope, 
the spot-size of the laser beam is $\sim 30\mu m$ in diameter and the position of the 
laser spot on the sample was fixed with an accuracy of $\pm 2.5\mu m$. The power of the 
incident laser light was set to 1.7 mW.

\section{Results and discussion}

\subsection{Optical microscopy}
The top panel of figure \ref{microscope}
shows microscope images of a Y$_{0.63}$Ca$_{0.37}$TiO$_3$ single crystal at $T=10$ K
in three different orientations, \ie facing a (001), (010), and (100) surface of the 
material, respectively, with respect to the orthorhombic Pbnm symmetry. For each of 
these different orientations, specific patterns of bright and dark regions appear 
which show a preferred direction. As a result, the whole sample surface partitions 
in stripe-like patterns with a width of the order of 10 - 20 $\mu$m. While on the 
(100)-surface, these stripes extend across the whole field of view of several 
hundreds of $\mu$m, on the other two surfaces they form disconnected islands 
of one kind with extensions of the order of 100 - 200 $\mu$m within a background 
of the other kind. The bottom panel of figure \ref{microscope} shows the evolution 
of the pattern with temperature from 100 K to room temperature  for the (001)-surface.
Comparing figure \ref{microscope}\subref{sampleC} and 
figure \ref{microscope}\subref{sampleC_100}, it is seen, that islands of brighter areas 
exist at $10$ K, which grow upon heating, while their shape is mainly maintained 
even at $100$ K. Upon further heating a growth of these islands in $a$ and $b$ 
directions of the sample continues. However, this  growth starts to take place 
more rapidly, so that these islands start to connect in $a$ and $b$ directions 
at $\sim 140$ K. Furthermore, at $\sim 150$ K, bright stripes evolving in $b$ 
direction start to occur within the dark background. The width of these 
sub-stripes is $\sim5\,\mu$m and the distance between the stripes is $\sim20\,\mu$m.
Finally, the brighter areas consisting of the former islands and the evolving stripes 
continue to grow and form a connected network as it is shown in 
figure \ref{microscope}\subref{sampleC_170} for $170$ K.

The contrast between bright and dark areas continuously weakens above $100$ K 
and eventually becomes insignificant at room temperature. However, by 
artificially enhancing the contrast of the microscope images, the different 
areas can be shown to still exist at 296 K, 
as shown in figure \ref{microscope}\subref{sampleC_296}.

A noteworthy observation is the fact that the bright areas invariably show up 
in the same positions after several temperature cycles. Given the exceptionally 
high doping level, it seems likely to assume chemical inhomogeneity of the 
material to be the reason for this pattern. Additionally, the microscope 
images show temperature hysteresis of the surface fractions in the same 
temperature region as reported in the resistivity data. Remarkable is, 
however, the unusually large length scales in the range of many tens to 
hundreds of $\mu$m, \ie several orders of magnitude larger than what had 
previously been reported.\cite{Matsuhata2004}

\subsection{Spatially resolved photoelectron spectroscopy}

\begin{figure}[htb]
	\includegraphics[width=.45\textwidth]{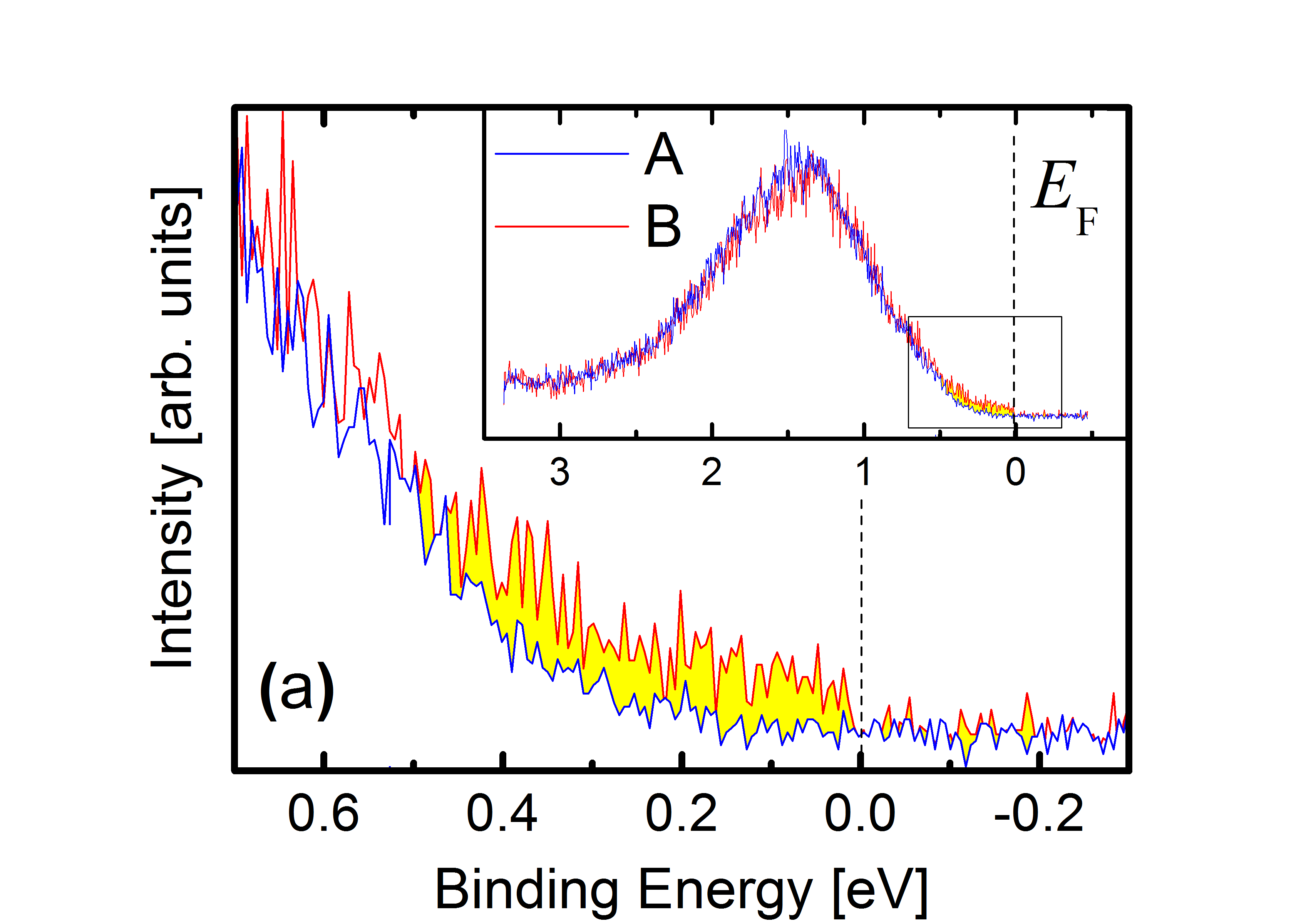}\\
	\includegraphics[width=.45\textwidth]{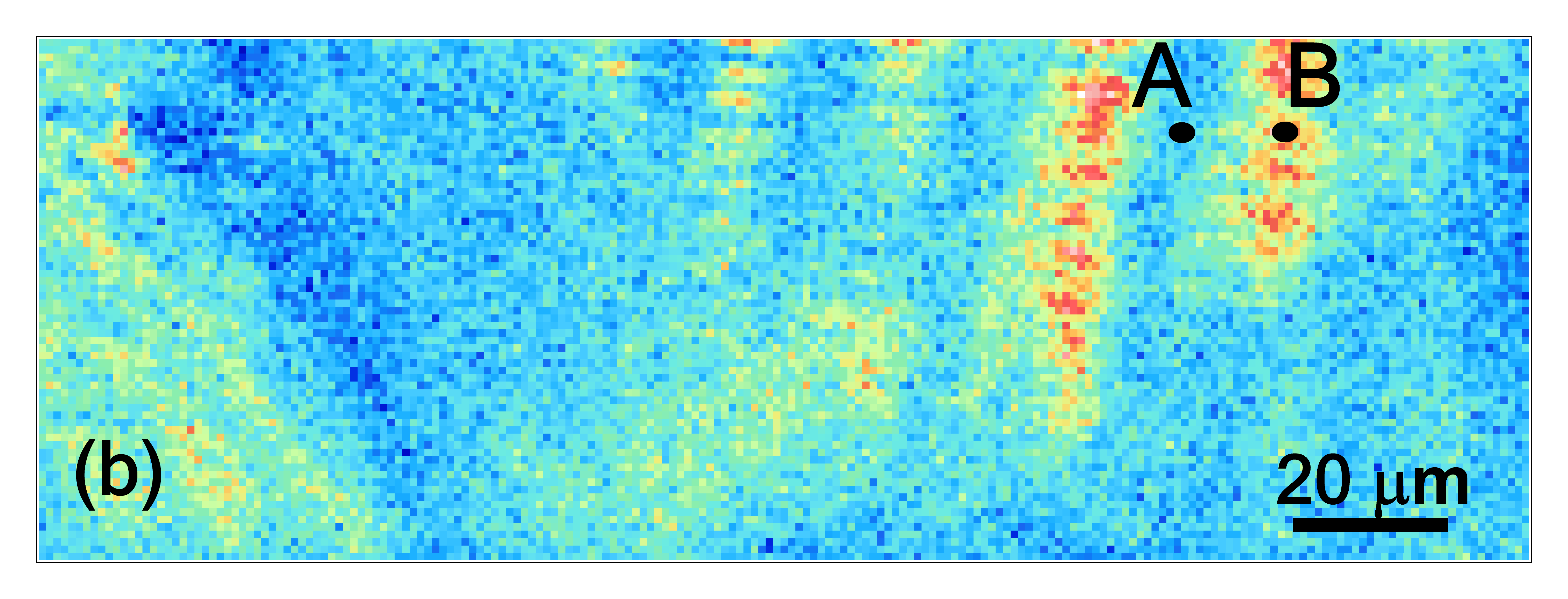}
	\captionsetup{justification=raggedright,singlelinecheck=false}
	\caption{(Color online) Spatially resolved photoemission spectra 
		of the Ti 3d band at 110K with sample orientation in [001]-direction. 
		(a) Comparison between conducting and insulating spectrum in red and blue, 
		respectively, at two different spots on the sample, denoted as 'A' and 'B'. 
		(b) $200\,\mu\text{m} \times 70\,\mu\text{m}$ cut-out of the spatially 
		resolved and normalized Ti 3d band photoemission spectra. Side 'A' and 'B' 
		are indicated by black dots.
	}
	\label{spatial_pes}
\end{figure}

In order to investigate the electronic properties of the different areas observed 
in the microscope images, spatially resolved photoelectron spectroscopy measurements 
have been carried out at $T=110$ K.
The inset in figure \ref{spatial_pes}a) shows the Ti 3$d$ spectral weight for two 
different sites on the sample, denoted as 'A' and 'B'. As the blowup in the main 
panel reveals, the spectra especially differ in a narrow region close to the Fermi 
level, with 'B' being slightly more conducting than 'A'.

Figure \ref{spatial_pes}b) shows a map of a $200\,\mu\text{m} \times 70\,\mu\text{m}$ 
area of the sample with a pixel size of $1\,\mu\text{m}^2$. The sample surface was 
oriented in [001]-direction and fractured before the measurement to get a clean surface. 
For each pixel, a full spectrum in the energy range as shown in the inset 
of figure \ref{spatial_pes}a) has been taken. Photoemission intensity variations 
due to surface roughness \etc have been corrected for by normalizing each 
spectrum to unity. The map shows the integrated intensity in a narrow energy 
window around the Fermi level, \ie the region where the spectra from 
sites 'A' and 'B' vary, as color values. Thus, more conducting regions 
appear bright and red in the figure in contrast to the insulating blue ones. 
The map reveals a similar domain pattern of stripes with a separation 
of $\approx 20\,\mu$m as had been observed in the microscope images. 
Furthermore, a smaller scale stripe formation is recognizable. 
The widths and distances between these sub-stripes  
are of the order of $\sim1\,\mu$m.
Stripe-like features are also observed in the photoemission spectral weights of 
the O $2p$ band which is a natural consequence of the hybridization with the Ti $3d$, 
and presumably related to a rigid band shift. The presumption based 
on the microscope observations, that the origin of the stripe pattern might be 
chemical inhomogeneity, appears to be in line with the results of the photoelectron spectroscopy.

\subsection{Spontaneous micro-Raman spectroscopy}

\begin{table}[htb]
\begin{tabular}{ccccc}
	\hline\hline
	Mode&  YMnO$_3$&  YMnO$_3$ &  YTiO$_3$& Y$_{0.63}$Ca$_{0.37}$TiO$_3$\\
	& Calc. & Exp.& Exp.& LTM - phase \\
	\hline
	A$_{g}$(7)&  104&  151&  145&  \\
	
	A$_{g}$(5)&  147&  188&  168&  \\
	
	A$_{g}$(2)&  223&  288&  273&  \\

	A$_{g}$(6)&  304&  323&  314&  \textcolor{blue}{294}\\
	
	A$_{g}$(4)&  407&  396&  417&  \textcolor{blue}{394}\\
	
	A$_{g}$(3)&  466&  497&  446&  \textcolor{blue}{457}\\
	
	A$_{g}$(1)&  524&  518&  512&  \textcolor{blue}{532}\\
\\
	B$_{1g}$(7)&  137&  151&  142&  \\
	
	B$_{1g}$(5)&  162&  220&  219&  \\
	
	B$_{1g}$(4)&  285&  317&  306&  \\
	
	B$_{1g}$(6)&  393&  341&  328&  \\
	
	B$_{1g}$(3)&  470&  481&  487&  \\
	
	B$_{1g}$(2)&  583&  537&  521&  \\
	
	B$_{1g}$(1)&  617&  616&  643&  \\
\\
	B$_{2g}$(5)&  145&  178&  151&  \\
	
	B$_{2g}$(4)&  363&  336&  284&  \\
	
	B$_{2g}$(3)&  390&  &  366&  \textcolor{blue}{342}\\
	
	B$_{2g}$(2)&  476&  &  452&  \textcolor{blue}{494}\\
	
	B$_{2g}$(1)	&  610&  &  678&  \textcolor{blue}{637}\\
\\
	B$_{3g}$(5)	&  181&  205&  162&  \\
	
	B$_{3g}$(3)	&  288&  284&  303&  \\
	
	B$_{3g}$(4)	&  342&  384&  &  \\
	
	B$_{3g}$(2)	&  413&  &  485&  \\
	
	B$_{3g}$(1)	&  593&  &  644&  \textcolor{blue}{593}\\
	\hline
	\hline
\end{tabular}
\captionsetup{justification=raggedright,singlelinecheck=false}
	\caption{Calculated and experimentally obtained phonon modes of YMnO$_{3}$,\cite{Iliev1998} YTiO$_{3}$,\cite{Tsurui2004} and the LTM phase of Y$_{0.63}$Ca$_{0.37}$TiO$_3$ 
		measured at 10 K in $c(aa)\bar{c}$ geometry, respectively. 
		The Raman shifts are given in units of cm$^{-1}$.}
	\label{modes}
\end{table}
Various Raman investigations on Y$_{1-x}$Ca$_{x}$TiO$_3$ are available in the literature, 
showing that the Raman spectra are quite different between insulating (x $<$ 0.33) and 
metallic (x $>$ 0.39) samples at low temperatures.\cite{Katsufuji1994,Tsurui2004}
However, the phonon energies at room temperature do not change significantly by doping.
This suggests, that mode assignments on YTiO$_3$ 
can be transferred to our doped material.\cite{Tsurui2004} Phonon mode calculations 
are available for the perovskitelike YMnO$_3$.\cite{Iliev1998}  
As the Ti$^{3+}$/Mn$^{3+}$ ions do not participate in the Raman active modes, 
no significant shift in the mode energies due to their different masses are expected. 
Table \ref{modes} shows the phonon energies for YMnO$_3$, both lattice dynamical 
calculations (LDC) and experimentally observed values compared to the experimentally 
obtained values for YTiO$_3$.\cite{Tsurui2004} This shows that there is indeed a 
good overall agreement in energies of both materials. Finally, the 
Raman shifts of Y$_{0.63}$Ca$_{0.37}$TiO$_3$ obtained in this work from fits of 
the Raman spectra are also listed in Tab. \ref{modes}. This comparison allows 
an explicit mode assignment of the observed phonon peaks.
The Raman active modes for Y$_{0.63}$Ca$_{0.37}$TiO$_3$  in Pbnm symmetry are 
assumed to be $\Gamma = 7A_g + 5B_{1g} + 7B_{2g} + 5B_{3g}$.
The Wyckoff positions are 4(c) for both Y and O1, 4(a) for Ti, and 8(d) for O2. 
But only Y, O1 and O2 yield Raman active modes.
The Raman scattered light corresponding to A$_g$-modes retains the polarization 
of the incoming light, refereed to in experiment as \emph{parallel polarization}, 
while for the B$_{1g}$-,B$_{2g}$-, and B$_{3g}$-modes, the polarization changes 
with the Raman process, thus these are observed in \emph{cross polarized} geometry.
The phase transition from Pbnm to P2$_1$/n preserves the total number of modes, 
because the primitive cell is the same in both phases and inversion symmetry is 
not broken. However, this transition leads to a redistribution of orthorhombic
B$_g$-modes to monoclinic A$_g$ ones. Therefore, the polarization dependence of 
the Raman modes can be used to identify the crystallographic phase of the material.  
\begin{figure*}[htbp]
	\includegraphics[width=.3\textwidth]{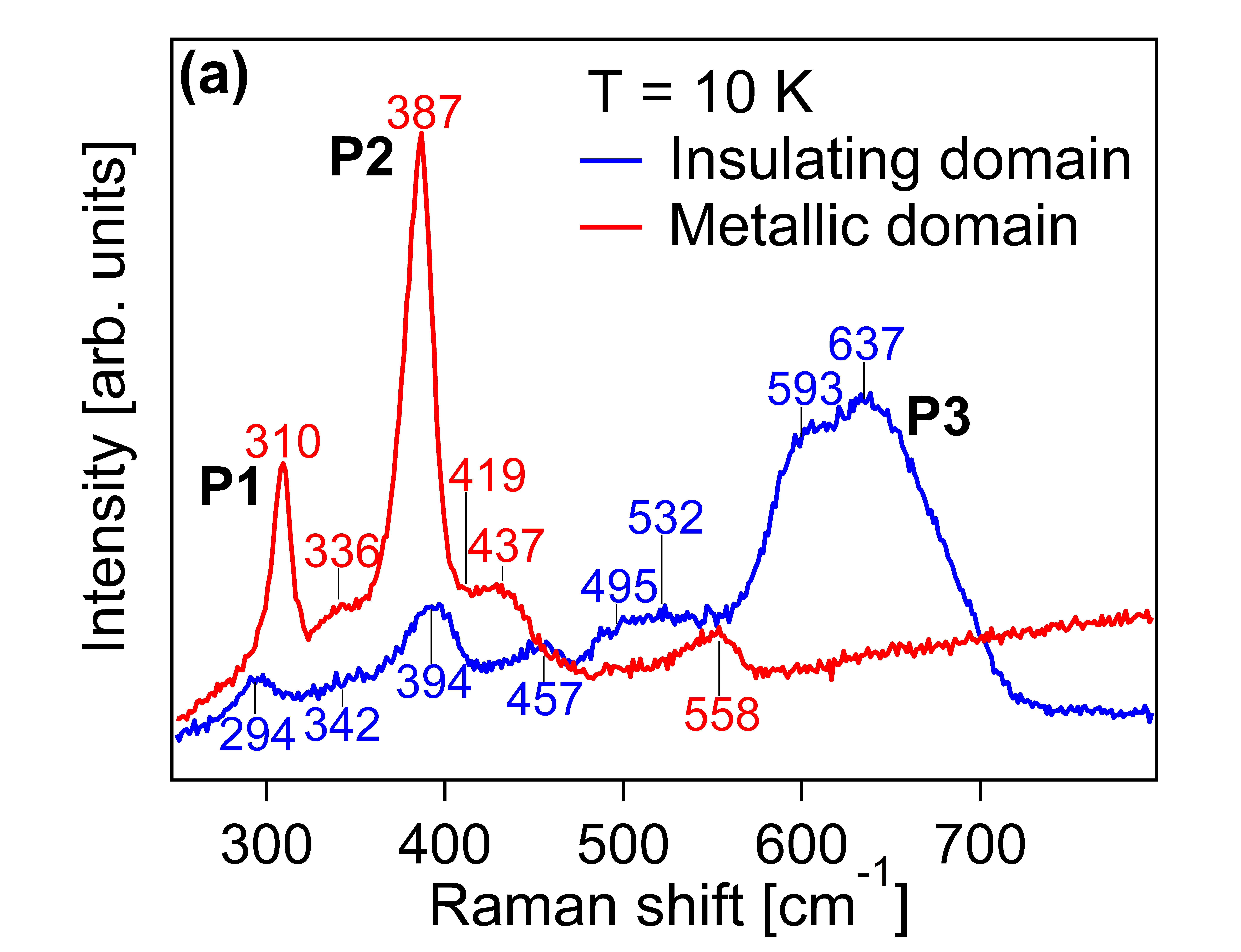}
	\includegraphics[width=.3\textwidth]{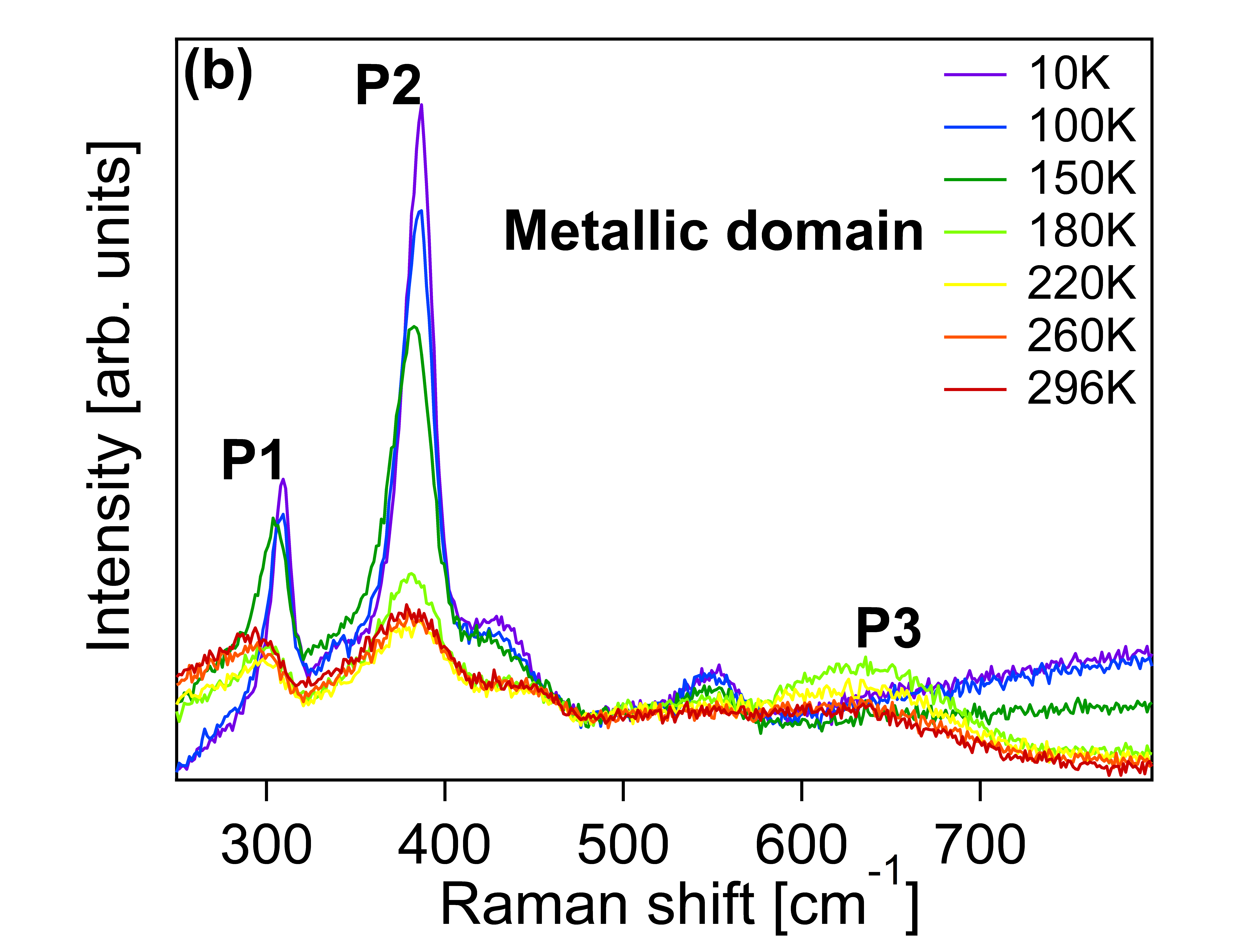}
	\includegraphics[width=.3\textwidth]{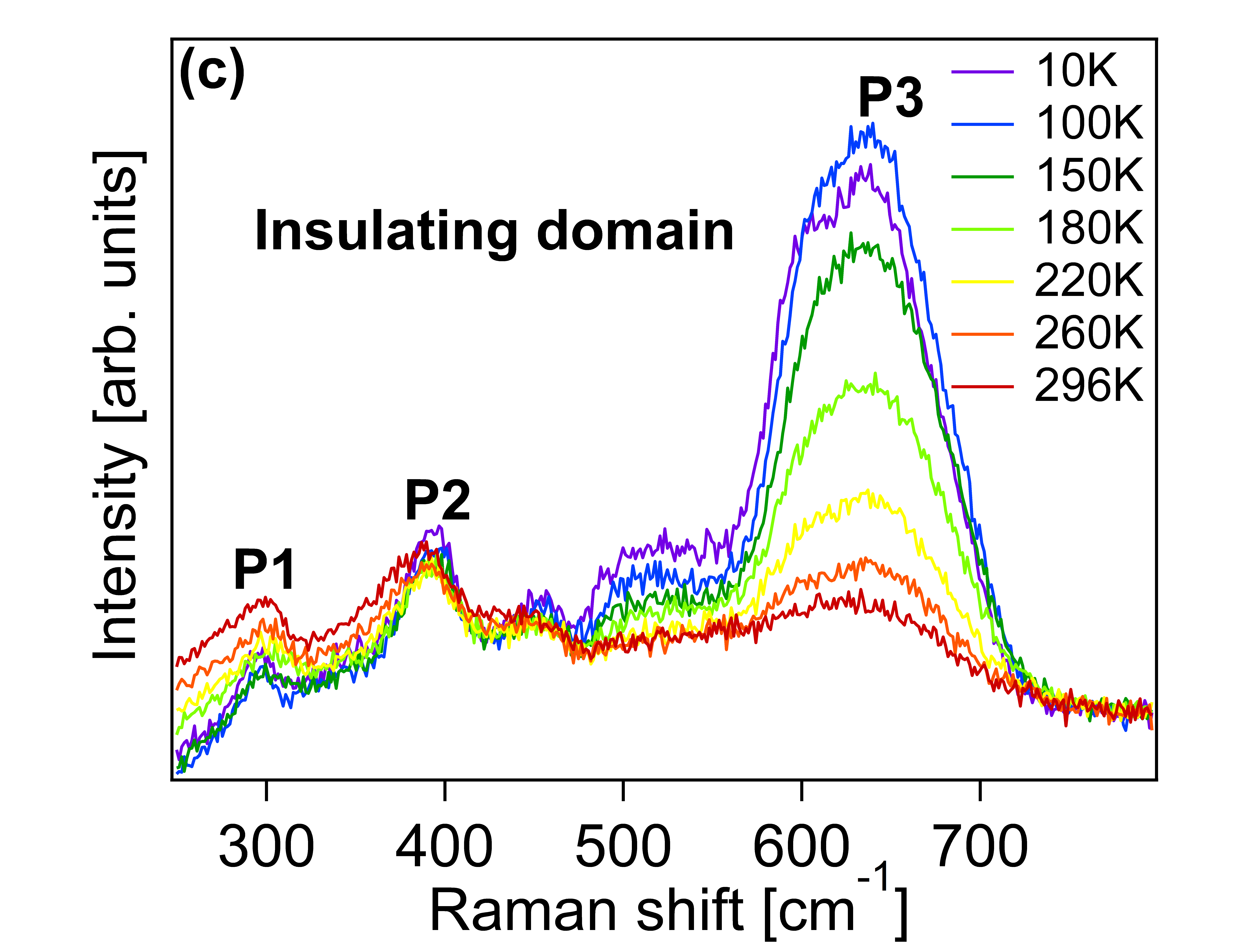}
	\captionsetup{justification=raggedright,singlelinecheck=false}
	\caption{(Color online)(a) Raman scattering spectra of \ycto single crystal,
		 measured for dark and bright spots, shown in figure \ref{sampleC}, 
		 in $c(aa)\bar{c}$ geometry  at 10 K. The spectrum in blue is obtained from a bright area, 
		 while the spectrum in red is obtained from a dark area. 
		 (b) and (c) temperature evolution of the Raman spectra of the dark and bright domains, respectively.}
	\label{raman_ramp}
\end{figure*}
\begin{figure}[htbp]
	\includegraphics[width=.35\textwidth]{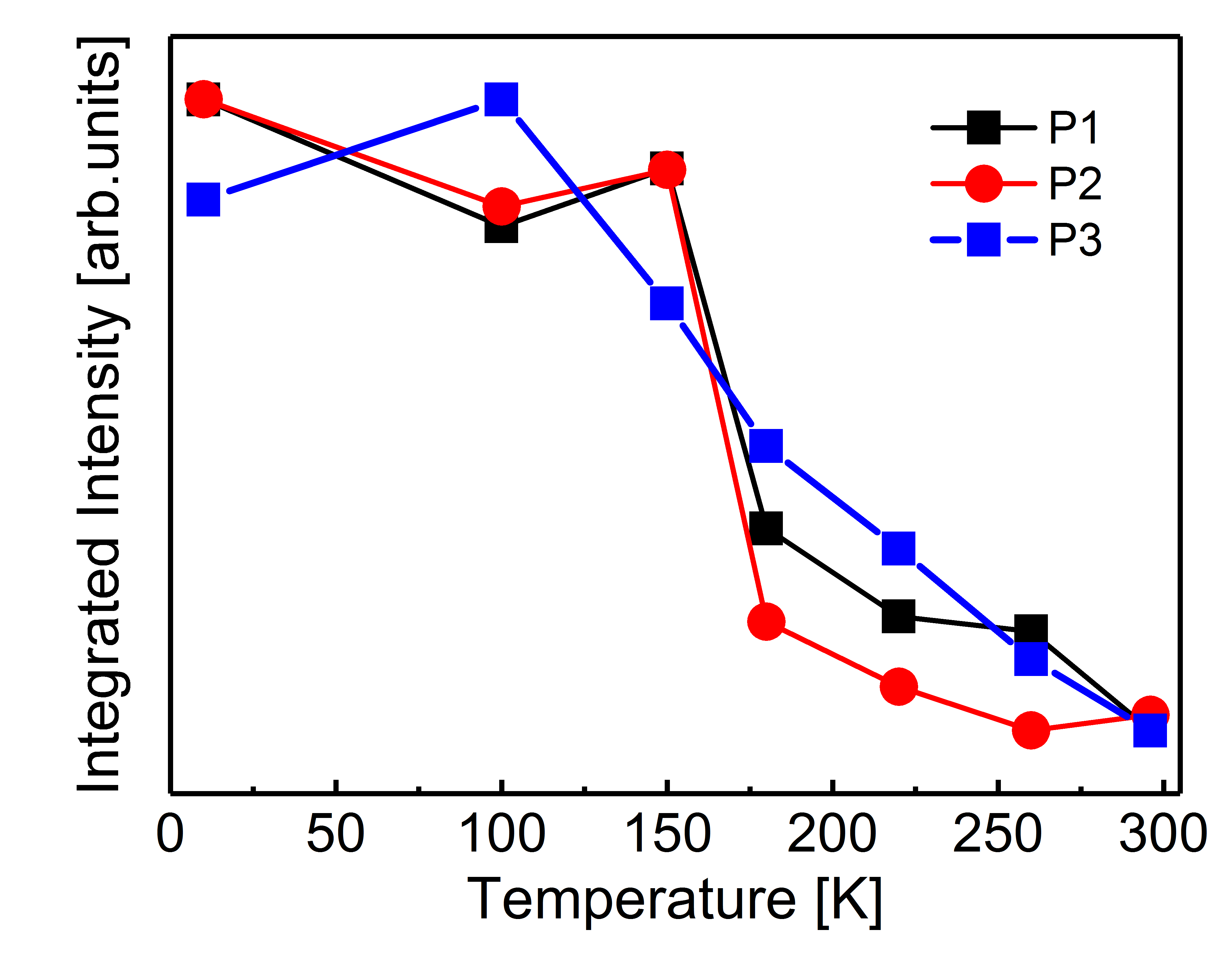}
	\captionsetup{justification=raggedright,singlelinecheck=false}
	\caption{(Color online) Temperature evolution  of the normalized integrated peak weights of 
		the  metallic (red and black)  and insulating (blue) domains, 
		represented by the peaks P1, P2, and P3, respectively.}
	\label{P1P2P3}
\end{figure}
Figure \ref{raman_ramp}a) shows the Raman spectra in c(aa)$\bar{\text{c}}$ geometry 
at 10 K, obtained from two different spots on the sample, corresponding to distinct phases, 
as described above in the optical microscopy section.
Significant differences are visible in the spectral weights and peak positions of 
the Raman spectra obtained from the bright and dark phases, respectively, see Fig.\ref{sampleC}.
The energy regions at $\sim$310, $\sim$390 and $\sim$640 cm$^{-1}$, 
denoted as P1, P2, and P3, respectively, exhibit the most prominent 
differences between the two phases.
As reported in Tsurui et $al.$ (2004),\cite{Tsurui2004}
the modes P1 and P2 are fingerprints of Raman spectra obtained from 
conducting samples with doping level $x > 0.39$, while P3 is only 
visible for doping levels $x < 0.39$, thus belonging to the insulating phase.  
Therefore, the Raman spectra allow us to identify the darker areas of the 
sample in Fig. \ref{sampleC}, as the LTO phase, represented by P1 and P2, 
and the brighter areas as the LTM phase, represented by P3.
Figure \ref{raman_ramp}b) and c) show the 
temperature dependence of the spectra for
spatially fixed positions in conducting and insulating regions, respectively, 
in the range from 10 K to room temperature.
Of particular interest is the feature P3, which 
is strongest in the insulating domain at 10 K. 
The polarization geometry and the calculated frequencies from the literature 
suggest that the feature consists of the B$_{2g}(1)$ breathing mode and of other 
bond-stretching modes. These become Raman active in parallel polarization in the 
monoclinic phase and show a strong temperature dependence.
Note, that the structural distortions, the modulation of the Ti-O bond distances, 
associated with orbital and charge ordering, are remarkably strong
in these titanates,\cite{KomarekPhD} which may explain the strong signal of the P3 
feature as well as the strong temperature dependence.

Following the temperature evolution of P3 from room temperature to 10 K, 
the peak weight continuously increases and saturates between 100 K and 150 K, 
see Figure \ref{P1P2P3}. Such a saturation behavior is also observed for the
static G-type distortion pattern associated with the charge and orbital 
order of the LTM phase.\cite{KomarekPhD} This similarity in saturation behavior indicates,
that the B$_{2g}(1)$ modes become A$_{1g}$ modes, which correspond to the static 
distortions of the LTM phase.
As the unit cell volumes of LTO and LTM phases differ, 
strain plays an important role in the phase separated state, 
which has been reported in TEM measurements:\cite{Matsuhata2004} it has been 
observed that especially the LTO phase is highly strained in contrast to the LTM phase,
which will favor phase separation.

The Raman spectra in the conducting domain show an abrupt change between 150 K and 180 K. 
This is especially seen in the peak weights of P1 and P2, 
shown as black and red curves in Figure \ref{P1P2P3}
Both modes require a description by an asymmetric Fano lineshape in the fits, which
may hint to an enhanced coupling of the charge carriers to the phonon modes P1 and P2.

At room temperature, the spectra in the different regions of the material are almost identical, 
in agreement with the observation that in the microscope images the 
contrast between them is barely detectable.
\section{Conclusions}
Combining our results from optical microscopy, 
spatially resolved photoelectron spectroscopy, and Raman scattering,
we can identify regions of different crystallographic and, in turn, 
electronic character in the material, which form stripe-like domain 
patterns of typical sizes of order $20\times100\,\mu\text{m}^2$. 
Along the direction of the stripes, these domains can extend up to several hundreds of $\mu$m.
The result of our investigations of Y$_{0.63}$Ca$_{0.37}$TiO$_3$ 
allow us to assign the bright domains, see Figure \ref{sampleC}, 
to be insulating and the dark domains to be conducting by using 
Raman spectroscopy. The insulating domains are furthermore identified 
with the LTM phase and the conducting domains with the LTO phase of the material. 
With these identifications, the optically observed percolation of the dark 
domains is indeed a percolation of the metallic LTO phase through the 
insulating LTM phase, which in the end leads to the macroscopic effect 
of a temperature-driven metal insulator transition. 
The domain patterns and the growth direction of the domains with 
varying temperature also explains the anisotropy observed in the resistivity.\cite{Tsurui2004}
As seen in the optical microscope pictures in Figure \ref{microscope}, 
the three differently oriented surfaces exhibit a preferred growth direction 
of the metallic domains in a- and c- directions. However the growth in 
b- direction is strongly suppressed. Resistivity measurements along a- and c- directions 
in the literature indeed show a temperature-driven metal to insulator transition, 
whereas in b- direction the MIT is strongly suppressed.\cite{Tsurui2004}
Different transition temperatures with varying stoichiometries (x=0.37-0.39) 
are reported in the literature.\cite{Kato2002,KomarekPhD} We attribute this to
a faster percolation due to the higher number of metallic domains, and vice versa.
In turn, the apparent natural inhomogeneity in the material is expected to result in 
spatially varying physical properties like transition temperatures.     
The reason for the stability of the insulating phase at high temperatures is still
unclear and remains an open question.

In summary, we have performed microstructural measurements 
using optical microscopy, spontaneous micro-Raman
scattering, and micro-photoemission
on single crystals of \ycto in a temperature range from 10 K to room temperature. 
We observe a phase separation throughout the temperature range and identify
the structural and the electronic differences between the 
different domains 
as well as their temperature dependence. Our results suggest
that the macroscopic change of the electronic transport behavior
is a result of
percolation of the conducting LTO phase
inside a non conducting phase, which at high temperature has an orthorhombic crystal structure, 
but at low temperature becomes monoclinic.

\section{ACKNOWLEDGMENTS}
We gratefully acknowledge fruitful discussions 
with C. Sch\"u\ss ler-Langeheine and D. Khomskii. 
This work has been supported by the Deutsche Forschungsgemeinschaft (DFG) 
through the Collaborative Research Center SFB 1238 (projects A02 and B03).


\begin{thebibliography}{18}
\expandafter\ifx\csname natexlab\endcsname\relax\def\natexlab#1{#1}\fi
\expandafter\ifx\csname bibnamefont\endcsname\relax
  \def\bibnamefont#1{#1}\fi
\expandafter\ifx\csname bibfnamefont\endcsname\relax
  \def\bibfnamefont#1{#1}\fi
\expandafter\ifx\csname citenamefont\endcsname\relax
  \def\citenamefont#1{#1}\fi
\expandafter\ifx\csname url\endcsname\relax
  \def\url#1{\texttt{#1}}\fi
\expandafter\ifx\csname urlprefix\endcsname\relax\def\urlprefix{URL }\fi
\providecommand{\bibinfo}[2]{#2}
\providecommand{\eprint}[2][]{\url{#2}}

\bibitem[{\citenamefont{Imada et~al.}(1998)\citenamefont{Imada, Fujimori, and
  Tokura}}]{Imada1998}
\bibinfo{author}{\bibfnamefont{M.}~\bibnamefont{Imada}},
  \bibinfo{author}{\bibfnamefont{A.}~\bibnamefont{Fujimori}}, \bibnamefont{and}
  \bibinfo{author}{\bibfnamefont{Y.}~\bibnamefont{Tokura}},
  \bibinfo{journal}{Rev. Mod. Phys.} \textbf{\bibinfo{volume}{70}},
  \bibinfo{pages}{1039} (\bibinfo{year}{1998}). 
  
  \bibitem[{\citenamefont{Yee}(2015)}]{Yee2015}
  \bibinfo{author}{\bibfnamefont{C.- H.}\bibnamefont{Yee}},
  \bibinfo{author}{\bibfnamefont{L.}\bibnamefont{Balents}},
  \bibinfo{journal}{Phys. Rev. X}, \textbf{\bibinfo{volume}{5}}
  \bibinfo{pages}{021007}, (\bibinfo{year}{2015}).
  
  \bibitem[{\citenamefont{Cao et~al.}(2015)\citenamefont{Cao, Shafer, Liu, Meyers, Kareev, Middey, Freeland, Arenholz, and Chakhalian}}]{Cao2015}
  \bibinfo{author}{\bibfnamefont{Y.} \bibnamefont{Cao}},
  \bibinfo{author}{\bibfnamefont{P.} \bibnamefont{Shafer}},
  \bibinfo{author}{\bibfnamefont{X.} \bibnamefont{Liu}},
  \bibinfo{author}{\bibfnamefont{D.} \bibnamefont{Meyers}},
  \bibinfo{author}{\bibfnamefont{M.} \bibnamefont{Kareev}},
  \bibinfo{author}{\bibfnamefont{S.}~\bibnamefont{Middey}},
  \bibinfo{author}{\bibfnamefont{J.~W.} \bibnamefont{Freeland}}, 
  \bibinfo{author}{\bibfnamefont{E.} \bibnamefont{Arenholz}},
  \bibnamefont{and}
  \bibinfo{author}{\bibfnamefont{J.} \bibnamefont{Chakhalian}},
  \bibinfo{journal}{Appl. Phys. Lett.} \textbf{\bibinfo{volume}{107}},
  \bibinfo{pages}{112401} (\bibinfo{year}{2015}).
  
\bibitem[{\citenamefont{Yang and Wu}(2017)}]{Yang2017}
  \bibinfo{author}{\bibfnamefont{X.}~\bibnamefont{Yang}} \bibnamefont{and}
  \bibinfo{author}{\bibfnamefont{G.}~\bibnamefont{Wu}},
  \bibinfo{journal}{EPL}, \textbf{\bibinfo{volume}{117}}
  \bibinfo{pages}{27004}, (\bibinfo{year}{2017}).
  
\bibitem[{\citenamefont{Katsufuji et~al.}(1995)\citenamefont{Katsufuji,
  Okimoto, and Tokura}}]{Katsufuji1995}
\bibinfo{author}{\bibfnamefont{T.}~\bibnamefont{Katsufuji}},
  \bibinfo{author}{\bibfnamefont{Y.}~\bibnamefont{Okimoto}}, \bibnamefont{and}
  \bibinfo{author}{\bibfnamefont{Y.}~\bibnamefont{Tokura}},
  \bibinfo{journal}{Phys. Rev. Lett.} \textbf{\bibinfo{volume}{75}},
  \bibinfo{pages}{3497} (\bibinfo{year}{1995}).

  \bibitem[{\citenamefont{Okimoto et~al.}(1995)}]{Okimoto1995}
\bibinfo{author}{\bibfnamefont{Y.}~\bibnamefont{Okimoto}}, 
\bibinfo{author}{\bibfnamefont{T.}~\bibnamefont{Katsufuji}},
\bibinfo{author}{\bibfnamefont{Y.}~\bibnamefont{Okada}},
\bibinfo{author}{\bibfnamefont{T.}~\bibnamefont{Arima}}, \bibnamefont{and}
\bibinfo{author}{\bibfnamefont{Y.}~\bibnamefont{Tokura}},
\bibinfo{journal}{Phys. Rev. B} \textbf{\bibinfo{volume}{51}},
\bibinfo{pages}{9581} (\bibinfo{year}{1995}).


\bibitem[{\citenamefont{Morikawa et~al.}(1996)\citenamefont{Morikawa, Mizokawa,
		Fujimori, Taguchi, and Tokura}}]{Morikawa1996}
\bibinfo{author}{\bibfnamefont{K.}~\bibnamefont{Morikawa}},
\bibinfo{author}{\bibfnamefont{T.}~\bibnamefont{Mizokawa}},
\bibinfo{author}{\bibfnamefont{A.}~\bibnamefont{Fujimori}},
\bibinfo{author}{\bibfnamefont{Y.}~\bibnamefont{Taguchi}} \bibnamefont{and}
\bibinfo{author}{\bibfnamefont{Y.}~\bibnamefont{Tokura}},
\bibinfo{journal}{Phys. Rev. B.}
\textbf{\bibinfo{volume}{54}}, \bibinfo{pages}{8446} (\bibinfo{year}{1996}).

\bibitem[{\citenamefont{Taguchi et~al.}(1993)\citenamefont{Taguchi, Tokura,
		Arima, and Inaba}}]{Taguchi1993}
\bibinfo{author}{\bibfnamefont{Y.}~\bibnamefont{Taguchi}},
\bibinfo{author}{\bibfnamefont{Y.}~\bibnamefont{Tokura}},
\bibinfo{author}{\bibfnamefont{T.}~\bibnamefont{Arima}}, \bibnamefont{and}
\bibinfo{author}{\bibfnamefont{F.}~\bibnamefont{Inaba}},
\bibinfo{journal}{Phys. Rev. B} \textbf{\bibinfo{volume}{48}},
\bibinfo{pages}{511} (\bibinfo{year}{1993}).


\bibitem[{\citenamefont{Tokura et~al.}(1993)\citenamefont{Tokura, Taguchi,
		Moritomo, Kumagai, Suzuki, and Iye}}]{Tokura1993}
\bibinfo{author}{\bibfnamefont{Y.}~\bibnamefont{Tokura}},
\bibinfo{author}{\bibfnamefont{Y.}~\bibnamefont{Taguchi}},
\bibinfo{author}{\bibfnamefont{Y.}~\bibnamefont{Moritomo}},
\bibinfo{author}{\bibfnamefont{K.}~\bibnamefont{Kumagai}},
\bibinfo{author}{\bibfnamefont{T.}~\bibnamefont{Suzuki}}, \bibnamefont{and}
\bibinfo{author}{\bibfnamefont{Y.}~\bibnamefont{Iye}},
\bibinfo{journal}{Phys. Rev. B} \textbf{\bibinfo{volume}{48}},
\bibinfo{pages}{14063} (\bibinfo{year}{1993}).



\bibitem[{\citenamefont{Iga et~al.}(1996)\citenamefont{Iga, Naka, Matsumoto,
		Shirakawa, Murata, and Nishihara}}]{Iga1996}
\bibinfo{author}{\bibfnamefont{F.}~\bibnamefont{Iga}},
\bibinfo{author}{\bibfnamefont{T.}~\bibnamefont{Naka}},
\bibinfo{author}{\bibfnamefont{T.}~\bibnamefont{Matsumoto}},
\bibinfo{author}{\bibfnamefont{N.}~\bibnamefont{Shirakawa}},
\bibinfo{author}{\bibfnamefont{K.}~\bibnamefont{Murata}}, \bibnamefont{and}
\bibinfo{author}{\bibfnamefont{Y.}~\bibnamefont{Nishihara}},
\bibinfo{journal}{Physica B}
\textbf{\bibinfo{volume}{223 \& 224}}, \bibinfo{pages}{526-528}
(\bibinfo{year}{1996}).

\bibitem[{\citenamefont{Kato et~al.}(2002)\citenamefont{Kato, Nishibori,
		Takata, Sakata, Nakano, Uchihira, Tsubota, Iga, and Takabatake}}]{Kato2002}
\bibinfo{author}{\bibfnamefont{K.}~\bibnamefont{Kato}},
\bibinfo{author}{\bibfnamefont{E.}~\bibnamefont{Nishibori}},
\bibinfo{author}{\bibfnamefont{M.}~\bibnamefont{Takata}},
\bibinfo{author}{\bibfnamefont{M.}~\bibnamefont{Sakata}},
\bibinfo{author}{\bibfnamefont{T.}~\bibnamefont{Nakano}},
\bibinfo{author}{\bibfnamefont{K.}~\bibnamefont{Uchihira}},
\bibinfo{author}{\bibfnamefont{M.}~\bibnamefont{Tsubota}},
\bibinfo{author}{\bibfnamefont{F.}~\bibnamefont{Iga}}, \bibnamefont{and}
\bibinfo{author}{\bibfnamefont{T.}~\bibnamefont{Takabatake}},
\bibinfo{journal}{J. Phys. Soc. Japan} \textbf{\bibinfo{volume}{71}},
\bibinfo{pages}{2082} (\bibinfo{year}{2002}).

  

\bibitem[{\citenamefont{Nakao et~al.}(2004)\citenamefont{Nakao, Tsubota, Iga,
		Uchihira, Nakano, Takabatake, Kato, and Murakami}}]{Nakao2004}
\bibinfo{author}{\bibfnamefont{H.}~\bibnamefont{Nakao}},
\bibinfo{author}{\bibfnamefont{M.}~\bibnamefont{Tsubota}},
\bibinfo{author}{\bibfnamefont{F.}~\bibnamefont{Iga}},
\bibinfo{author}{\bibfnamefont{K.}~\bibnamefont{Uchihira}},
\bibinfo{author}{\bibfnamefont{T.}~\bibnamefont{Nakano}},
\bibinfo{author}{\bibfnamefont{T.}~\bibnamefont{Takabatake}},
\bibinfo{author}{\bibfnamefont{K.}~\bibnamefont{Kato}}, \bibnamefont{and}
\bibinfo{author}{\bibfnamefont{Y.}~\bibnamefont{Murakami}},
\bibinfo{journal}{J. Phys. Soc. Japan} \textbf{\bibinfo{volume}{73}},
\bibinfo{pages}{2620} (\bibinfo{year}{2004}).




\bibitem[{\citenamefont{Tsubota et~al.}(2003)\citenamefont{Tsubota, Iga,
		Nakano, Uchihira, Kura, Takemura, Bando, Umeo, Takabatake, Nishibori
		et~al.}}]{Tsubota2003}
\bibinfo{author}{\bibfnamefont{M.}~\bibnamefont{Tsubota}},
\bibinfo{author}{\bibfnamefont{F.}~\bibnamefont{Iga}},
\bibinfo{author}{\bibfnamefont{T.}~\bibnamefont{Nakano}},
\bibinfo{author}{\bibfnamefont{K.}~\bibnamefont{Uchihira}},
\bibinfo{author}{\bibfnamefont{S.}~\bibnamefont{Kura}},
\bibinfo{author}{\bibfnamefont{M.}~\bibnamefont{Takemura}},
\bibinfo{author}{\bibfnamefont{Y.}~\bibnamefont{Bando}},
\bibinfo{author}{\bibfnamefont{K.}~\bibnamefont{Umeo}},
\bibinfo{author}{\bibfnamefont{T.}~\bibnamefont{Takabatake}},
\bibinfo{author}{\bibfnamefont{E.}~\bibnamefont{Nishibori}},
\bibinfo{author}{\bibfnamefont{M.}~\bibnamefont{Takata}},
\bibinfo{author}{\bibfnamefont{M.}~\bibnamefont{Sakata}},
\bibinfo{author}{\bibfnamefont{K.}~\bibnamefont{Kato}}, \bibnamefont{and}
\bibinfo{author}{\bibfnamefont{Y.}~\bibnamefont{Ohishi}}, \bibinfo{journal}{J. Phys. Soc. Japan}
\textbf{\bibinfo{volume}{72}}, \bibinfo{pages}{3182} (\bibinfo{year}{2003}).



\bibitem[{\citenamefont{Matsuhata et~al.}(2004)\citenamefont{Matsuhata, Iga,
		Tsubota, Nakano, Takabatake, and Kato}}]{Matsuhata2004}
\bibinfo{author}{\bibfnamefont{H.}~\bibnamefont{Matsuhata}},
\bibinfo{author}{\bibfnamefont{F.}~\bibnamefont{Iga}},
\bibinfo{author}{\bibfnamefont{M.}~\bibnamefont{Tsubota}},
\bibinfo{author}{\bibfnamefont{T.}~\bibnamefont{Nakano}},
\bibinfo{author}{\bibfnamefont{T.}~\bibnamefont{Takabatake}},
\bibnamefont{and} \bibinfo{author}{\bibfnamefont{K.}~\bibnamefont{Kato}},
\bibinfo{journal}{Phys. Rev. B}
\textbf{\bibinfo{volume}{70}}, \bibinfo{pages}{134109} (\bibinfo{year}{2004}).


\bibitem[{\citenamefont{Komarek}(2009)}]{KomarekPhD}
\bibinfo{author}{\bibfnamefont{A. C.}~\bibnamefont{Komarek}}, Dissertation,
\bibinfo{school}{Univ. Cologne} (\bibinfo{year}{2009}), https://kups.ub.uni-koeln.de/2982/


	\bibitem[{\citenamefont{Tsubota}(2005)}]{Tsubota2005}
	\bibinfo{author}{\bibfnamefont{M.}~\bibnamefont{Tsubota}},
	\bibinfo{author}{\bibfnamefont{F.}~\bibnamefont{Iga}},
	\bibinfo{author}{\bibfnamefont{K.}~\bibnamefont{Uchihira}},
	\bibinfo{author}{\bibfnamefont{T.}~\bibnamefont{Nakano}},
	\bibinfo{author}{\bibfnamefont{S.}~\bibnamefont{Kura}},
	\bibinfo{author}{\bibfnamefont{M.}~\bibnamefont{Takabatake}},
	\bibinfo{author}{\bibfnamefont{S.}~\bibnamefont{Kodoma}},
	\bibinfo{author}{\bibfnamefont{H.}~\bibnamefont{Nakao}},
	\bibnamefont{and} \bibinfo{author}{\bibfnamefont{Y.}~\bibnamefont{Murakami}},
	\bibinfo{journal}{J. Phys. Soc. Jpn.} \textbf{\bibinfo{volume}{74}}, 12, \bibinfo{pages}{3259-3266} (\bibinfo{year}{2005}) 

\bibitem[{\citenamefont{Barbo et~al.}(2000)}]{Barbo2000}
\bibinfo{author}{\bibfnamefont{F.}~\bibnamefont{Barbo}}, 
\bibinfo{author}{\bibfnamefont{M.}~\bibnamefont{Bertolo}},
\bibinfo{author}{\bibfnamefont{A.}~\bibnamefont{Bianco}},
\bibinfo{author}{\bibfnamefont{G.}~\bibnamefont{Cautero}},
\bibinfo{author}{\bibfnamefont{S.}~\bibnamefont{Fontana}},
\bibinfo{author}{\bibfnamefont{T.K.}~\bibnamefont{Johal}},
\bibinfo{author}{\bibfnamefont{S.}~\bibnamefont{La Rosa}},
\bibinfo{author}{\bibfnamefont{K.}~\bibnamefont{Kaznacheyev}},
\bibnamefont{and}
  \bibinfo{author}{\bibfnamefont{G.}~\bibnamefont{Margaritondo}},
  \bibinfo{journal}{Rev. Sci. Instrum.} \textbf{\bibinfo{volume}{71}},
  \bibinfo{number}{1} (\bibinfo{year}{2000}).


\bibitem[{\citenamefont{Katsufuji and Tokura}(1994)}]{Katsufuji1994}
\bibinfo{author}{\bibfnamefont{T.}~\bibnamefont{Katsufuji}} \bibnamefont{and}
\bibinfo{author}{\bibfnamefont{Y.}~\bibnamefont{Tokura}},
\bibinfo{journal}{Phys. Rev. B} \textbf{\bibinfo{volume}{50}},
\bibinfo{pages}{2704} (\bibinfo{year}{1994}).


\bibitem[{\citenamefont{Tsurui et~al.}(2004)\citenamefont{Tsurui, Ogita,
  Udagawa, Tsubota, and Iga}}]{Tsurui2004}
\bibinfo{author}{\bibfnamefont{T.}~\bibnamefont{Tsurui}},
  \bibinfo{author}{\bibfnamefont{N.}~\bibnamefont{Ogita}},
  \bibinfo{author}{\bibfnamefont{M.}~\bibnamefont{Udagawa}},
  \bibinfo{author}{\bibfnamefont{M.}~\bibnamefont{Tsubota}}, \bibnamefont{and}
  \bibinfo{author}{\bibfnamefont{F.}~\bibnamefont{Iga}},
  \bibinfo{journal}{Phys. Rev. B} \textbf{\bibinfo{volume}{69}},
  \bibinfo{pages}{024102} (\bibinfo{year}{2004}).
  
  
\bibitem[{\citenamefont{Iliev et~al.}(1998)\citenamefont{Iliev, Lee, Popov,
  Sun, Thomsen, Meng, and Chu}}]{Iliev1998}
\bibinfo{author}{\bibfnamefont{M.~N.} \bibnamefont{Iliev}},
  \bibinfo{author}{\bibfnamefont{M.~V.} \bibnamefont{Abrashev}},
  \bibinfo{author}{\bibfnamefont{H.-G.} \bibnamefont{Lee}},
  \bibinfo{author}{\bibfnamefont{V.~N.} \bibnamefont{Popov}},
  \bibinfo{author}{\bibfnamefont{Y.~Y.} \bibnamefont{Sun}},
  \bibinfo{author}{\bibfnamefont{C.}~\bibnamefont{Thomsen}},
  \bibinfo{author}{\bibfnamefont{R.~L.} \bibnamefont{Meng}}, \bibnamefont{and}
  \bibinfo{author}{\bibfnamefont{C.~W.} \bibnamefont{Chu}},
  \bibinfo{journal}{Phys. Rev. B} \textbf{\bibinfo{volume}{57}},
  \bibinfo{pages}{2872} (\bibinfo{year}{1998}).



\end{thebibliography}
\end{document}